\documentstyle[prl,psfig,
aps]{revtex}
\twocolumn

\begin{document}
\draft
\twocolumn[\hsize\textwidth\columnwidth\hsize\csname @twocolumnfalse\endcsname

\title{ Tight-binding parameters and exchange integrals of \\
Ba$_2$Cu$_3$O$_4$Cl$_2$ }

\author{H. Rosner, R. Hayn}
\address{ Max Planck Arbeitsgruppe "Elektronensysteme",
Technische Universit\"at Dresden, \\
D-01062 Dresden, Federal Republic of Germany }

\author{ J. Schulenburg}
\address{ Institut f\"ur Theoretische Physik, Universit\"at Magdeburg, \\
D-39016 Magdeburg, Federal Republic of Germany }

\date{\today}

\maketitle

\begin{abstract}
Band structure calculations for Ba$_2$Cu$_3$O$_4$Cl$_2$ within the
local density approximation (LDA) are presented.
The investigated compound is similar to the antiferromagnetic parent
compounds of cuprate superconductors but contains additional Cu$_B$
atoms in the planes. Within, the LDA metallic behavior is found with
two bands crossing the Fermi surface (FS).  These bands are built
mainly from Cu $3d_{x^2-y^2}$ and O $2p_{x,y}$ orbitals, and a
corresponding tight-binding (TB) model has been parameterized. All
orbitals can be subdivided in two sets corresponding to the A- and
B-subsystems, respectively, the coupling between which is found to be
small. To describe the experimentally observed antiferromagnetic
insulating state, we propose an extended Hubbard model with the
derived TB parameters and local correlation terms characteristic for
cuprates.
Using the derived parameter set we calculate the
exchange integrals for the Cu$_3$O$_4$ plane. The results are in
quite reasonable agreement with the experimental values for the
isostructural compound Sr$_2$Cu$_3$O$_4$Cl$_2$.
\end{abstract}
\pacs{71.10.+x,75.50.Ee,75.30.Et}
]
Several new cuprate materials were studied in the last years
which have modifications in the standard CuO$_2$ plane. That class
includes not only the now famous ladder compounds Sr$_{n-1}$Cu$_{n+1}$O$_{2n}$
\cite{dagotto96}, but also Ba$_2$Cu$_3$O$_4$Cl$_2$ or the similar
Sr$_2$Cu$_3$O$_4$Cl$_2$ \cite{kipka76,noro94,ohta95,yamada95}. The
latter two compounds are antiferromagnetic insulators which are
characterized by a Cu$_3$O$_4$ plane with additional Cu$_B$ atoms in
the standard Cu$_A$O$_2$ plane (Figs.\ 1 and 3). This new class of
cuprates is currently of large scientific interest since it may shed
light on some of the open questions in high-$T_c$ superconductivity. 
The layered
cuprate Ba$_2$Cu$_3$O$_4$Cl$_2$ itself is also interesting due to
its rich magnetic structure and the unconventional character of its
lowest energy excitations. Experimentally, two N\'eel temperatures have been
found $T_N^A \sim 330$ K and $T_N^B \sim 31$ K
\cite{noro94,yamaguchi97} connected with the two sublattices of A- and
B-copper. The magnetic susceptibility and the small ferromagnetic
moment have been explained phenomenologically \cite{chou97} together
with a determination of the exchange integrals. Like in undoped
Sr$_2$CuO$_2$Cl$_2$ \cite{wells95}, the lowest electron removal states in
Ba$_2$Cu$_3$O$_4$Cl$_2$ can be interpreted in terms of Zhang-Rice
singlets \cite{zhang88} with a new branch of singlet excitations
connected with the B-sublattice \cite{golden97} .

\par
One might expect that the additional Cu$_B$ give rise to considerable
differences in the 
electronic structure in comparison with the usual CuO$_2$
plane. In particular, the amount of coupling between both subsystems seems to be
crucial. That 
question shall be addressed here by a microscopic investigation starting from a
bandstructure calculation. However, as for most undoped cuprates, the local
density approximation (LDA) leads to a metallic behavior and one has to treat
the electron correlation in a more explicit way. In this paper, we use 
bandstructure calculations to determine tight-binding (TB) parameters and then
add the local Coulomb correlation terms. In contrast to the preliminary
TB fit \cite{rosner97} we now distinguish between different oxygen
orbitals. Furthermore, 
we  estimate the exchange integrals. 

\par {\bf A. Bandstructure calculation.} The tetragonal unit cell of
Ba$_2$Cu$_3$O$_4$Cl$_2$ is shown in Fig.\ 1.  
We performed
LDA calculations for this substance using the linear combination of
atomic orbitals (LCAO).
Due to the relatively open structure 8 empty spheres per unit cell
(between Cu$_A$ and Cl) have been introduced.  The calculation was
scalar relativistic and we have chosen a minimal basis consisting of
Cu(4$s$,4$p$,3$d$), O(2$s$,2$p$), Ba(6$s$,6$p$,5$d$) and
Cl(3$s$,3$p$). To optimize the local basis a contraction potential has
been used at each site \cite{eschriglcao}. The Coulomb part of the
potential was constructed as a sum of overlapping spherical
contributions and the exchange and correlation part was treated in the
atomic sphere approximation. The resulting bandstructure is shown in
Fig.\ 2a. Its main result are two bands crossing the Fermi surface
(FS). These two bands, and most other bands also, have nearly no
dispersion in z-direction which justifies the restriction 
to the Cu$_3$O$_4$ plane  in the following discussion.
Analyzing the orbital
character of the two relevant bands we found that the broad band is
built up of predominantly the Cu$_A$ 3$d_{x^2-y^2}$ orbital which
hybridizes with one part of the plane oxygen 2$p_{x,y}$ orbitals
resulting in  $dp\sigma$ bonds. The corresponding oxygen orbitals, directed
to the Cu$_A$ atoms, will be denoted here as $p$ orbitals. They are to
be distinguished from the oxygen $\pi$ orbitals which are
perpendicular to them \cite{mattheiss89} (see Fig.\ 3). Those oxygen
$\pi$ orbitals hybridize with Cu$_B$ 3$d_{x^2-y^2}$ building the
narrow band at the FS. So we have to consider 11 orbitals in the
elementary cell of Cu$_3$O$_4$, namely 2 Cu$_A$ 3$d_{x^2-y^2}$, 1
Cu$_B$ 3$d_{x^2-y^2}$, 4 oxygen $p$ and 4 oxygen $\pi$ orbitals. In
Fig.\ 2b we pick out the corresponding bands from the
LDA-bandstructure for which the sum of all 11 orbital projections is
large. It is seen that the two bands crossing the FS have a nearly pure
3$d_{x^2-y^2}$ and 2$p_{x,y}$ character \cite{remark1}.  
It is important to note that the sum of all orbital projections in Fig.\ 2b
decreases with increasing binding energy. That can be explained since
for larger binding energy more weight goes into the overlap density in
our LCAO which does not contribute to Fig.\ 2b. One can also observe
that the lower 8 bands in Fig.\ 2b are not as pure as the upper
three. For the upper bands only a very small weight of additional
orbitals, in particular Cu 4 $s$ contributions, has been detected. These
contributions are neglected in the following.  
\par {\bf B. Tight-binding
parameters.} It is our main goal to find a TB description of the
relevant bands crossing the FS. This task is difficult due to the
large number of bands between -1 eV and -8 eV of Fig.\ 2a.  There is
no isolated band complex which makes a TB analysis easy. But we have
pointed out already that the relevant bands are a nearly pure
combination of Cu $d_{x^2-y^2}$ and O $p_{x,y}$ orbitals. So we will
only concentrate on these orbitals, accepting some deviations in the
lower band complex between \mbox{-3} and -8 eV. The relevant orbitals
are depicted in Fig.\ 3.
We distinguish two classes of
orbitals, one consists of Cu$_A$ 3$d_{x^2-y^2}$ orbitals with on-site
energies $\varepsilon_d^A$ and oxygen $p$ orbitals with
$\varepsilon_p$, the other class incorporates Cu$_B$ 3$d_{x^2-y^2}$
with on-site energies $\varepsilon_d^B$ and oxygen $\pi$ orbitals with
$\varepsilon_{\pi}$.  The coupling between both classes of orbitals
which correspond to the Cu$_A$- and Cu$_B$-subsytems, respectively, is
provided by the parameter $t_{p \pi}$. Corresponding to the different
orbitals, one has to distinguish 4 on-site energies
($\varepsilon_d^A$, $\varepsilon_d^B$, $\varepsilon_p$,
$\varepsilon_{\pi}$), the nearest neighbor transfer integrals
($t_{pd}$, $t_{\pi d}$), and several kinds of oxygen transfers
($t_{pp}$, $t_{\pi\pi}^1$, $t_{\pi\pi}^2$, $t_{p\pi}$), all together
10 parameters, which are sketched in Fig.\ 3.  

We found that the Cu$_A$-Cu$_B$ transfer $t_{dd}$ can be neglected since its
estimation yields a value smaller than 0.08 eV \cite{remark2}.  Each
$p$ orbital is located between two Cu$_A$ sites and we neglect the
influence of different local environments on the $t_{pp}$ transfer
integral. In the case of the $t_{\pi\pi}$ transfer we distinguish the
two possible local arrangements, but the numerical difference between
$t_{\pi\pi}^1$ and $t_{\pi\pi}^2$ is small (see Table \ref{TBM}). The
necessity to distinguish between oxygen $p$- and $\pi$-orbitals was
first pointed out by Mattheiss and Hamann \cite{mattheiss89} for
the case of the standard CuO$_2$ plane.

\par 
Since there is a considerable admixture of other orbitals,
especially Cu 3$d_{xy}$ and Cu 3$d_{3z^2-r^2}$, in some of the lower
bands of Fig.\ 2b, we cannot determine the 10 TB parameters by a least
square fit of the 11 TB bands to the heavily shaded LDA bands  of
Fig.\ 2b. Instead, at the high symmetry points $\Gamma=(0,0)$ and
$M=(\pi/a,\pi/a)$ we picked out those bands in Fig.\ 2b which have the
most pure 3$d_{x^2-y^2}$ and 2$p_{x,y}$ character. Only those energies
were compared with the TB bandstructure (Fig.\ 2c) derived by
diagonalizing a $11 \times 11$ matrix.  In this way it is possible to
calculate the parameter set analytically because the TB matrix splits
up into $3 \times 3$ and $4 \times 4$ matrices at the high symmetry
points $\Gamma=(0,0)$ and $M=(\pi/a,\pi/a)$.  The resulting parameters
are given in Table \ref{TBM}.  These values are similar to that which
are known for the standard CuO$_2$ plane. The largest transfer
integrals are $t_{pd}=1.43$ eV and $t_{\pi d}=1.19$ eV as
expected. But they are somewhat smaller than in the previous TB fit
\cite{rosner97} where all oxygen orbitals have been treated to be
identical. The difference between $t_{pp}$ and $t_{\pi\pi}$ is roughly
a factor of 2 in coincidence with the situation in the standard
CuO$_2$ plane \cite{mattheiss89}. We have found that only the smallest
parameter, $t_{p\pi}=0.25$ eV, is responsible for the coupling between
the subsystems of Cu$_A$ and Cu$_B$. Thus despite the fact that the two
oxygen $p$ and $\pi$ orbitals are located in real space at the same
atom, they are quite far away from each other in the Hilbert space.

\par
{\bf C. Exchange integrals.} Thus far we have found that
the TB parameters are rather similar to the standard CuO$_2$ case and that
the coupling between Cu$_A$- and Cu$_B$-subsystem is quite small. This
justifies the usage of standard parameters for the Coulomb interaction
part of the Hamiltonian. Of course it would be desirable to determine 
these values by a
constrained density functional calculation for Ba$_2$Cu$_3$O$_4$Cl$_2$, but we
expect only small changes in the estimation of exchange integrals
presented below.

\par The Coulomb interaction also changes the on-site copper and
oxygen energies. Their difference, given in Table \ref{TBM}, is too
small to explain the charge transfer gap of $\sim 2$ eV in
Ba$_2$Cu$_3$O$_4$Cl$_2$ \cite{schmelz97}.  Adding 2 eV to the on-site
oxygen energies, the difference $\Delta=\varepsilon_p -
\varepsilon_d^A$ (in hole representation which is chosen from now on)
becomes similar to the standard value derived by Hybertsen {\em et
al.}\ \cite{hybertsen90} for La$_2$CuO$_4$. We have used the values of
Ref.\ \cite{hybertsen90} also for $U_d$, $U_p$, $U_{pd}$ and
$K_{pd}$. Since we now have two oxygen orbitals at one site we also
have to take into account the corresponding Hund's rule coupling
energy. Unfortunately, that correlation energy is known with less
accuracy than the other one and we choose here the simple rule
$J_H^O=-0.1 U_p$. The Coulomb repulsion between two oxygen holes in
$p$- and $\pi$-orbitals is assumed to be $U_{p\pi}=U_p+2J_H^O$, which
is a valid approximation given degenerate orbitals.  In the second
line of Table \ref{TBM} we combine the TB parameters derived from the
bandstructure of Ba$_2$Cu$_3$O$_4$Cl$_2$ (now in hole representation)
with the standard Coulomb correlation terms. This parameter set then
defines a 11 band extended Hubbard model for the Cu$_3$O$_4$ plane
which is used for the following estimation.
 
\par The exchange integrals have been calculated using the usual
Rayleigh-Schr\"odinger perturbation theory on small clusters (Fig.\
4).  All transfer integrals (and $J_H^O$) have been considered as a
perturbation around the local limit. We calculated all exchange
integrals in the corresponding lowest order. The exchange
$J_{AA}\propto t_{pd}^4/\Delta^3$ between two Cu$_A$ spins is given in
4th order for the simple Cu$_A$-O-Cu$_A$ cluster. It turns out that
the influence of intersite Coulomb and exchange terms $U_{pd}$ and
$K_{pd}$ is rather large, decreasing $J_{AA}$ from 246 meV to 99 meV
(see Table \ref{JMOD}).  In spite of our rather approximate procedure,
the latter value agrees quite reasonably with the phenomenological
value (130 $\pm$ 40) meV \cite{chou97} for
Sr$_2$Cu$_3$O$_4$Cl$_2$. $J_{AA}$ is thus also quite close to the
standard value of the CuO$_2$ plane ($\sim 140$ meV
\cite{hybertsen90}).

\par
The exchange $J_{BB}$ is given only in 6th order for a larger cluster of two
Cu$_B$, one Cu$_A$ and 4 oxygen orbitals. Correspondingly, it is roughly one
order of magnitude smaller, $J_{BB}\sim 17$ meV (Table \ref{JMOD}). For $J_{AB}$ we
need to distinguish antiferromagnetic and ferromagnetic contributions. There
are two AFM couplings between nearest neighbor copper atoms $J_{AB,af}^{(1)}$
and third nearest neighbor copper atoms $J_{AB,af}^{(3)}$, both being
comparably small at 4.6 and 0.8 meV, respectively. The ferromagnetic
contribution $J_{AB,f}$= -20meV between nearest neighbor copper spins arises in 5th
order and is provided by Hund's rule coupling of two virtual oxygen holes
sitting at the same oxygen. Due to the uncertainty in $J_H^O$, this
value has to be taken with care. 

\par
Summarizing, we presented a LCAO-LDA bandstructure calculation for
Ba$_2$Cu$_3$O$_4$Cl$_2$. Deriving TB parameters from it, we found only a weak
coupling between the two sets of orbitals connected with the subsystems of
copper A (Cu$_A$ $d_{x^2-y^2}$ and O $p$) and copper B (Cu$_B$ $d_{x^2-y^2}$
and O $\pi$). Furthermore, we found exchange integrals $J_{AA}$, $J_{BB}$ 
and $J_{AB}$ in reasonable agreement \cite{remark3} with
phenomenologically derived values from
magnetic susceptibility data if we add to the TB parameters the standard local
Coulomb correlation energies. 
\par
The authors thank S.-L.\ Drechsler, H.\ Eschrig, J.\ Richter, J.\ Fink, M.S.\
Golden, H.\ Schmelz and A.\ 
Aharony for useful discussions.

\cleardoublepage
\newpage
\parindent0cm
\begin{center}
\begin {minipage}{16cm}
\begin{table}\label{TBM}
\begin{tabular}{ c c c c c c c c c c c c c c c c c }
parameter&$\varepsilon_d^A$&$\varepsilon_d^B$&$\varepsilon_p$&$\varepsilon_\pi$&$t_{pd}$&$t_{\pi
d}$&$t_{pp}$&$t_{\pi\pi}^1$&$t_{\pi\pi}^2$&$t_{p\pi}$&$U_d$&$U_p$&$U_{pd}$&$U_{p\pi}$&$K_{pd}$&$J_H^O$\\
\hline \hline
TB fit &&&&&&&&&&&&&&&&\\
(electron representation)&-2.50&-2.12&-4.68&-3.73&1.43&1.19&0.81&0.41&0.50&0.25&-&-&-&-&-&-\\
(energy/eV) &&&&&&&&&&&&&&&&\\
\hline
extended Hubbard model &&&&&&&&&&&&&&&&\\
(hole representation)&2.50&2.12&6.68&5.73&-1.43&-1.19&-0.81&-0.41&-0.50&-0.25&10.5&4.0&1.2&3.2&-0.18&-0.4\\
(energy/eV) &&&&&&&&&&&&&&&&\\
\end{tabular}
\caption{}
\end{table}
\end {minipage}
\par
\end{center}

\parindent0cm
\vspace{0.30cm}
\begin{center}
\begin{table}\label{JMOD}
\begin{tabular}{ c c c c }
exchange & without $U_{pd},K_{pd}$&with $U_{pd},K_{pd}$&experiment\\
integral &(meV) &(meV) &(meV) \\
\hline \hline
$J_{AA}$         & 246 & 99 & 130 $\pm$ 40\\ 
$J_{BB}$         & 26  & 17 & 10  $\pm$ 1 \\ 
$J_{AB,af}^{(1)}$& 6.9 & 4.6&      -      \\ 
$J_{AB,f}^{(1)}$   &     &-20 &      -      \\ 	
$J_{AB}^{(1)}$     &     &-15 &-12  $\pm$ 9 \\ 
$J_{AB,af}^{(3)}$     & 1.4 & 0.8&      -      \\ 
\end{tabular}
\caption{}
\end{table}
\end{center}

\newpage
\cleardoublepage

\figure
{FIG.\ 1\
The body centered tetragonal unit cell of Ba$_2$Cu$_3$O$_4$Cl$_2$
with lattice constants a=5.51 \AA\ and c=13.82 \AA\ .
}
\figure
{FIG.\ 2\
{\bf{a)}} LCAO-LDA band structure in the Cu$_3$O$_4$ plane of
Ba$_2$Cu$_3$O$_4$Cl$_2$, the Fermi level is at zero energy.
{\bf{b)}} The same as in a), but the weight of the lines is scaled
with the sum of all 11 orbital projections that are used in the TB model.
{\bf{c)}} The band structure of the TB model. The parameter set used is
shown in Table \ref{TBM}.
The wavevector is measured in units of ($\pi/a$,$\pi/a$).
}
\figure
{FIG.\ 3\ 
Two elementary cells of the Cu$_3$O$_4$ plane in
Ba$_2$Cu$_3$O$_4$Cl$_2$ or Sr$_2$Cu$_3$O$_4$Cl$_2$  
and the Cu $3d_{x^2-y^2}$ and O $2p_{x,y}$
orbitals comprising the TB model. 
Also shown are the corresponding transfer integrals
$t_{pd}$, $t_{\pi d}$, $t_{pp}$, $t_{\pi\pi}^1$, $t_{\pi\pi}^2$ and
$t_{p\pi}$.  The Cu$_A$ orbitals with onsite energy $\varepsilon_d^A$
are marked by black diamonds, the Cu$_B$ orbitals with onsite energy
$\varepsilon_d^B$ by black squares and the two different kinds of O
orbitals with onsite energies $\varepsilon_p$ and $\varepsilon_\pi$,
respectively, by black circles. The orbitals of the B-subsystem are
shaded to distinguish them from the orbitals of the A-subsystem
(white).
}
\figure
{FIG.\ 3\
Clusters used for the calculation of the exchange integrals {\bf{a)}}
$J_{AA}$, {\bf{b)}}  $J_{BB}$, {\bf{c)}} $J_{AB,af}^{(1)}$ and
$J_{AB,f}^{(1)}$, {\bf{d)}}
$J_{AB,af}^{(3)}$.  The Cu$_A$ sites are marked with {\bf{A}}, the
Cu$_B$ sites with {\bf{B}} and the oxygen sites with {\bf{O}}.
}
\table
{TABLE I\
{
Parameters of the TB fit and the proposed extended Hubbard
model for the Cu$_3$O$_4$ plane in Ba$_2$Cu$_3$O$_4$Cl$_2$.
}
\table
{TABLE II\
{
Different exchange integrals as explained in the
text. Compared are estimations within the extended Hubbard model with
experimental values \protect\cite{chou97}.
}


\begin{thebibliography}{10}

\bibitem{dagotto96}
E. Dagotto and T.~M. Rice, Science {\bf 271},  618  (1996).

\bibitem{kipka76}
R. Kipka and Hk. M\"uller-Buschbaum, Z. Anorg. Allg. Chem. {\bf 419},  58  (1976).

\bibitem{noro94}
S. Noro {\it et~al.}, Materials Science and Engineering {\bf 25},  167  (1994).

\bibitem{ohta95}
H. Ohta {\it et~al.}, J. Phys. Soc. Jpn. {\bf 64},  1759  (1995).

\bibitem{yamada95}
K. Yamada, N. Suzuki, and J. Akimitsu, Physica (Amsterdam) {\bf 213\&\ 214B},  191
   (1995).

\bibitem{yamaguchi97}
T. Ito, H. Yamaguchi, and K. Oka, Phys. Rev. B {\bf 55},  R684  (1997).

\bibitem{chou97}
F.~C. Chou {\it et~al.}, Phys. Rev. Lett. {\bf 78},  535  (1997).

\bibitem{wells95}
B.~O. Wells {\it et~al.}, Phys. Rev. Lett. {\bf 74},  964  (1995).

\bibitem{zhang88}
F.~C. Zhang and T.~M. Rice, Phys. Rev. B {\bf 37},  3759  (1988).

\bibitem{golden97}
M.~S. Golden {\it et~al.}, Phys. Rev. Lett. {(to be published)}.

\bibitem{rosner97}
H. Rosner and R. Hayn, Physica B {(to be published)}.

\bibitem{eschriglcao}
H. Eschrig, {\em {Optimized LCAO Method}}, 1. ed. (Springer-Verlag, Berlin,
  1989).

\bibitem{mattheiss89}
L.~F. Mattheiss and D.~R. Hamann, Phys. Rev. B {\bf 40},  2217  (1989).

\bibitem{remark1}
{The corresponding band complex with nearly pure 3d$_{x^2-y^2}$ and 2p$_{x,y}$
  character includes also a third band just below the FS. It has Cu$_A$
  3$d_{x^2-y^2}$ and O $p$ character similar to the broad band crossing the FS.
  This can be understood since there are two Cu$_A$ in the elementary cell of
  Cu$_3$O$_4$}.

\bibitem{remark2}
{The parameter $t_{dd}$ can be roughly estimated by the weight of the Cu$_A$ in
  the Cu$_B$ band crossing the Fermi-level at the $\Gamma=(0,0)$ point of the
  Brillouin zone. At this point the coupling via $t_{p\pi}$ is not possible due
  to symmetry.}

\bibitem{schmelz97}
H.~C. Schmelz {\it et~al.}, Physica B {(to be published)}.

\bibitem{hybertsen90}
M.~S. Hybertson, E. Stechel, M. Schl{\"u}ter, and D.~R. Jennison, Phys. Rev. B
  {\bf 41},  11068  (1990).

\bibitem{remark3}
{The anisotropic coupling $J \sim$ 20 $\mu$eV which was found to be responsible
  for the small ferromagnetic moment in Sr$_2$Cu$_3$O$_4$Cl$_2$ cannot be
  estimated within the  model proposed here. It requires a more refined
  treatment incorporating spin-orbit coupling and more orbitals at the Cu
  site.}

\end{thebibliography}
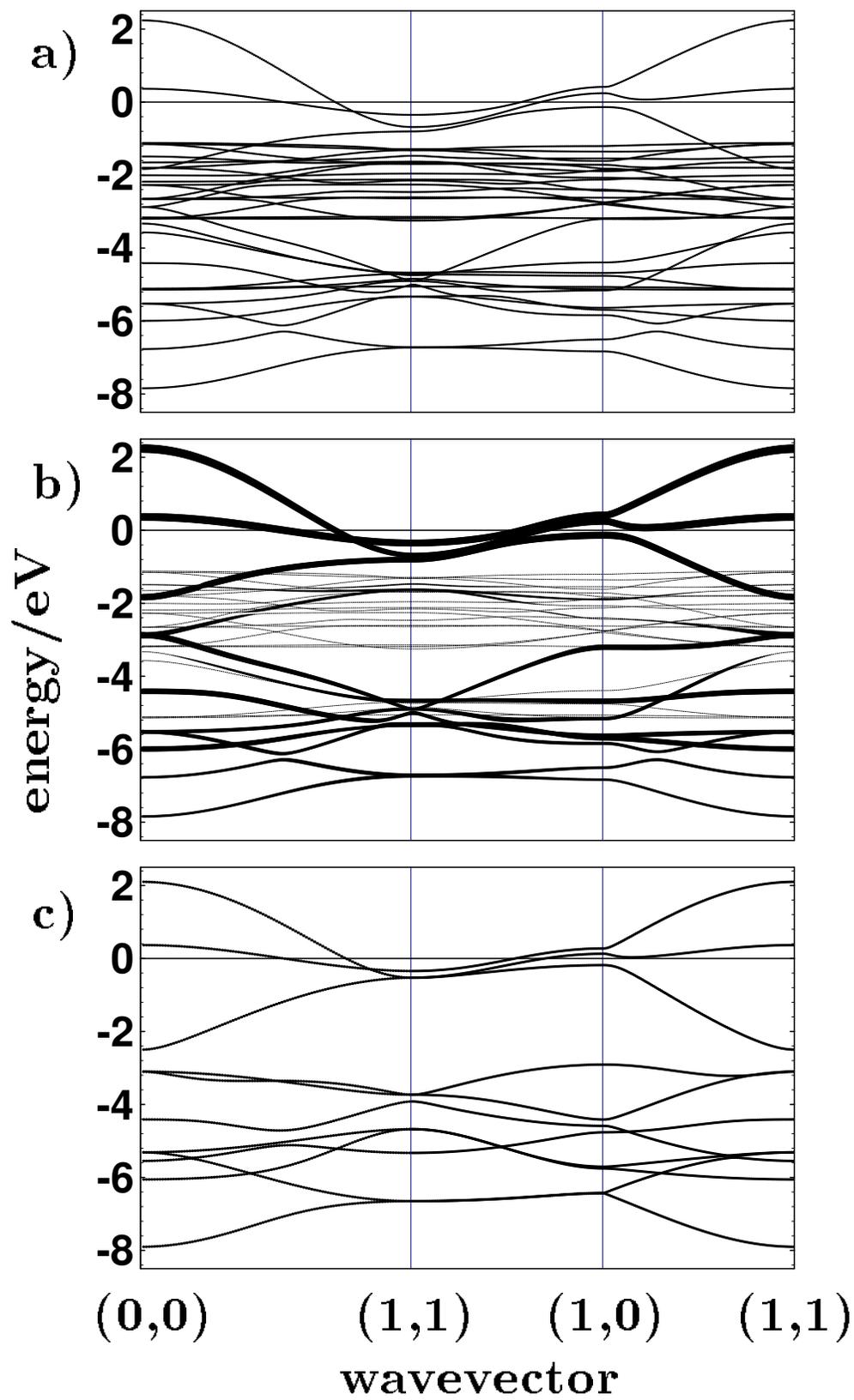
\end{document}